\documentclass[fleqn,10pt]{wlscirep}
\usepackage[utf8]{inputenc}
\usepackage[T1]{fontenc}
\usepackage{lineno}

\usepackage{longtable}
\usepackage{enumitem}
\usepackage{tablefootnote}
\usepackage{tabularray}
\usepackage{graphicx}
\usepackage{caption}
\usepackage[caption=false]{subfig}
\usepackage[export]{adjustbox}
\usepackage{tcolorbox}
\usepackage{float}
\floatstyle{plaintop}
\usepackage{multirow}

\definecolor{amber}{rgb}{1.0, 0.75, 0.0}

\title{CODE-ACCORD: A Corpus of building regulatory data for rule generation towards automatic compliance checking}

\author[1,$\dag$, *]{Hansi Hettiarachchi}
\author[2,$\dag$]{Amna Dridi}
\author[2]{Mohamed Medhat Gaber}
\author[2]{Pouyan Parsafard}
\author[2]{Nicoleta Bocaneala}
\author[3]{Katja Breitenfelder}
\author[4]{Gonçal Costa}
\author[5]{Maria Hedblom}
\author[6]{Mihaela Juganaru-Mathieu}
\author[7]{Thamer Mecharnia}
\author[3]{Sumee Park}
\author[5]{He Tan}
\author[2]{Abdel-Rahman H. Tawil}
\author[2]{Edlira Vakaj}

\affil[1]{Faculty of Science and Technology, Lancaster University, Lancaster LA1 4WA, UK}
\affil[2]{Faculty of Computing, Engineering and Built Environment, Birmingham City University, Birmingham B4 7XG, UK}
\affil[3]{Fraunhofer Institute for Building Physics IBP, Department Indoor Climate and Climatic Impacts, Fraunhofer Str. 10, 83626 Valley, Germany}
\affil[4]{Human Environment Research (HER), La Salle, Ramon Llull University, Barcelona,  Catalonia, Spain}
\affil[5]{Department of Computing, School of Engineering, Jönköping University, Box 1026, 551 11, Jönköping, Sweden}
\affil[6]{Mines Saint-Etienne, Institut Henri Fayol, Département ISI, F - 42023, Saint-Etienne, France}
\affil[7]{Université de Lorraine, CNRS, LORIA, 54506, Vandœuvre-lès-Nancy, France}

\affil[*]{corresponding author(s): Hansi Hettiarachchi (h.hettiarachchi@lancaster.ac.uk)}

\affil[$\dag$]{these authors contributed equally}

\begin{abstract}
Automatic Compliance Checking (ACC) within the Architecture, Engineering, and Construction (AEC) sector necessitates automating the interpretation of building regulations to achieve its full potential. Converting textual rules into machine-readable formats is challenging due to the complexities of natural language and the scarcity of resources for advanced Machine Learning (ML). Addressing these challenges, we introduce CODE-ACCORD, a dataset of 862 sentences from the building regulations of England and Finland. Only the self-contained sentences, which express complete rules without needing additional context, were considered as they are essential for ACC. Each sentence was manually annotated with entities and relations by a team of 12 annotators to facilitate machine-readable rule generation, followed by careful curation to ensure accuracy. The final dataset comprises 4,297 entities and 4,329 relations across various categories, serving as a robust ground truth. CODE-ACCORD supports a range of ML and Natural Language Processing (NLP) tasks, including text classification, entity recognition, and relation extraction. It enables applying recent trends, such as deep neural networks and large language models, to ACC. 
\end{abstract}
\begin{document}

\flushbottom
\maketitle

\thispagestyle{empty}


\section*{Background \& Summary}
Building codes establish regulations and standards dictating the minimum safety and welfare requirements for buildings and structures. Compliance with these codes throughout a building's lifecycle, including design, construction, and renovation/demolition, is essential to guarantee the stability, safety, usability and reliability of building designs. Although compliance checks have traditionally been conducted manually, there is now a growing motivation to automate this process due to the significant time and human resources demanded by the manual approach \cite{fuchs2021natural, zhang2023unpacking}. The advancement of more semantically rich Building Information Models (BIMs) has further encouraged this tendency, making Automated Compliance Checking (ACC) achievable \cite{MALSANE201551}. However, since building codes are written in text, as the first step, their underlying information needs to be extracted and converted into machine-readable formats to enable ACC via an intelligent approach \cite{zhou2022integrating}. However, extracting information from text has been a challenge due to the complexities associated with natural language, especially due to the unstructured nature and the human-centred design \cite{Zhang2012, Locatelli2021}.

Early efforts in machine-readable rule formalisation focused on developing software tools and predefined procedures that domain experts could use to format textual information into rules manually. One such approach involved RASE tagging \cite{hjelseth2011capturing}, which annotates text blocks into requirements (R), applicabilities (A), selections (S), and exceptions (E), together with metadata to help formulate rules \cite{Beach2015rule}. Another approach used an object-oriented procedure to filter declarative (or computer-interpretable) clauses and extract their physical entities and attributes, creating an elemental view of building regulations to facilitate easier information processing \cite{MALSANE201551}. Similarly, a logical rule-based mechanism using noun and verb phrases was introduced for Korean building codes \cite{LEE201649}. However, the manual conversion process was time-consuming and required the constant involvement of domain experts, prompting researchers to seek automated solutions.

Building on this trend, rule-based approaches were commonly used to extract information from regulatory text to generate machine-readable rules \cite{zhang2016semantic, zhang2017integrating, zhou2017ontology, xue2020building}. Both semantic and syntactic features have been used to design the rules, considering the complexities associated with the text. These methods were also integrated with varied natural language processing (NLP) techniques, such as syntactic parsing that analyses text structure using domain-specific context-free grammars (CFGs) to improve accuracy \cite{zheng2022knowledge, zhou2022integrating}. However, rule-based methods generally lack adaptability across different domains since most rules are specific to a particular domain \cite{zhang2021deep}. Also, designing rules requires intensive manual efforts, which are time-consuming and error-prone and demand extensive domain expertise. To mitigate these difficulties, there has been a shift towards supervised learning-based approaches, recently. Deep learning techniques, in particular, have shown great promise for extracting information from regulatory texts, aligning with the recent trends in NLP \cite{shen2022parallel, shen2023diffusionner, plum2022biographical, yang2023histred}. Various architectures such as Bidirectional Long Short-Term Memory (Bi-LSTM), Convolutional Neural Networks (CNN) and transformers have been adopted to extract information effectively \cite{zhang2021deep, wang2021deep, zheng2022pre, zhou2022integrating, wang2023deep}. More recently, this task has also been approached as a machine translation problem, converting text into formats like LegalRuleML \cite{athan2013oasis} using advanced text generation models such as T5 and BART \cite{fuchs2022neural}.

While supervised learning approaches overcome some challenges of earlier methods, such as reducing the need for extensive involvement of human or domain experts and saving time, they rely heavily on high-quality annotated data for training. Various strategies have been employed to annotate regulatory data and capture key information. For instance, one study introduced a dataset in JSON format aligned with an ontology for Chinese building codes \cite{zheng2022knowledge}. Another applied a similar approach to German building codes, using data-specific attributes defined by domain experts \cite{recski2024brise}. However, these datasets are often limited to specific domains (e.g. fire protection) or locations (e.g. Vienna), as they depend heavily on predefined information structures like ontologies. Such dependencies make the annotation strategies difficult to adapt to other domains or regions, requiring significant revisions to the underlying structures. Additionally, there is a noticeable lack of openly available datasets.

To address these limitations, this study aims to develop a simple annotation strategy that can be generalised across building codes from different sub-domains and regions. Our approach focuses specifically on extracting information from natural language, which poses challenges due to its unstructured nature. On closer examination, text typically consists of two fundamental types of information: (1) entities (also referred to as named entities) and (2) relations, which are essential for understanding the conveyed ideas \cite{Jurafsky2009speech}. An entity is a specific piece of information or a concept that can be categorised, or simply, anything that can be referred to using a proper name \cite{Jurafsky2009speech}. For example, given the sentence \textit{``A fire door must be self-closing.''}, \textit{``fire door''} and \textit{``self-closing''} describe entities. A relation is a semantic connection/association between an entity pair \cite{Jurafsky2009speech}. For example, there is a \textit{``necessity''} relation between the entities \textit{``fire door''} and \textit{``self-closing''} in the above sentence. Altogether, entities and relations form a network/knowledge graph that captures the rule(s) expressed in the text. Building on this concept, we propose an annotation strategy based on a generic set of entities and relations to convert unstructured text into machine-readable formats. Using this strategy, we annotated a dataset named CODE-ACCORD and made it publicly available to encourage the adoption of modern NLP and supervised learning techniques to advance the ACC processes.

In summary, the CODE-ACCORD corpus comprises 862 self-contained sentences extracted from the building regulations of England and Finland. A self-contained sentence is defined as a regulatory sentence that expresses a rule and contains all the details itself without any linguistical co-references that are unresolvable within the sentence, references to external sources or incomplete/ambiguous concepts. Such sentences are essential for ACC as they express rules that can be directly extracted and interpreted without extensive cross-referencing or additional context. Each sentence was manually annotated for entities and relations, with subsequent rounds of curation to ensure accuracy. Overall, CODE-ACCORD contains 4,297 annotated entities distributed across four categories and 4,329 relations distributed across ten categories.

\section*{Methods}

The development of the CODE-ACCORD corpus involved two main stages: (1) data collection and (2) data annotation. Initially, sentences were carefully extracted, and then, they underwent a thorough annotation process, resulting in the final dataset.

\subsection*{Data Collection Methodology} \label{sec:dataMethodology}

Our data collection approach mainly focused on extracting sentences that describe rules from building regulatory data to support our ultimate goal of creating a dataset that enables the automatic generation of machine-readable rules.

\subsubsection*{Data Sources}
CODE-ACCORD utilised the published building regulations of England and Finland, which are openly available, as its primary/raw data sources. England's building codes were collected from the official website of the UK Department for Levelling Up, Housing and Communities and the Ministry of Housing, Communities and Local Government (\url{https://www.gov.uk/government/collections/approved-documents}). Finland's building codes were collected from the official website of the National Building Code of Finland from the Ministry of Environment (\url{https://ym.fi/en/the-national-building-code-of-finland}).

The English translation of the Finnish National Building Code was used as this work aimed to build a corpus in English. England's building regulations included the guidelines and standards that dictate the construction and maintenance of buildings in England. These regulations were designed to ensure the safety, health, and welfare of people in and around buildings, as well as to conserve fuel and power in these structures. They cover various aspects of building construction, including structural integrity, fire safety, accessibility, energy efficiency, and more. These regulations are organised into different chapters or sections, each of which addresses a specific domain or aspect of construction, such as Part A: Structure, Part B: Fire Safety, Part K: Protection from Falling and Part L: Conservation of Fuel and Power. Finland's building regulations are similar to England's regulations in terms of their purpose but may have variations in specific requirements to suit the local context. They are issued as official government decrees organised into sections or chapters, such as Accessibility, Fire Safety and Energy Efficiency, each addressing specific domains or functional requirements. These regulations mainly guide the planning, design, construction, and maintenance of buildings to meet the country's standards for safety and environmental considerations. 

In both England and Finland, building regulations are published in PDF documents which are available online to the public. There is a combined total of $33$ documents, consisting of $23$ documents for English regulations, which span $1548$ pages, and 10 documents for Finnish regulations, covering $140$ pages. Table~\ref{tab:Description} presents the statistical overview of the data sources encompassing both English and Finnish regulations.

\begin{table}[h!]
\caption{Description and Statistics of CODE-ACCORD data sources}\label{tab:Description}%
\begin{tabular}{@{}llrr@{}}
\toprule
Regulations & Approved Document/Decree  & \#Volumes & \#Pages \\
\midrule
England  & A: Structure  & 1 & 54   \\
    & B: Fire Safety   & 2 & 384  \\
    & C: Site preparation and resistance to contaminates and moisture   & 1 & 52   \\
    & D: Toxic Substances   & 1 & 10   \\
   & E: Resistance to Sound   & 1 & 86   \\
   & F: Ventilation   & 2 & 110   \\
   & G: sanitation, hot water safety and water efficiency   & 1 & 55   \\
   & H: drainage and waste disposal   & 1 & 64   \\
   & J: Combustion appliances and fuel storage systems    & 1 & 89   \\
   & K: Protection from falling, collision and impact    & 1 & 68   \\
   & L: Conservation of fuel and power     & 2 & 220   \\
    & M: Access to and use of buildings      & 2 & 143   \\
      & O: Overheating      & 1 & 44   \\
      & P: Electrical safety       & 1 & 22   \\
      & Q: Security in dwellings       & 1 & 20   \\
       & R: Infrastructure for electronic communications      & 2 & 56   \\
          & S: Infrastructure for charging electric vehicles      & 1 & 47   \\
        & Material and workmanship: Approved Document 7      & 1 & 24   \\

    \hline
    Finland & Accessibility & 1 & 6 \\
            & Fire Safety & 1 & 25 \\
            & Energy Efficiency & 1  & 18 \\
            & Planning and Supervision & 1  & 7 \\
            & Strength and Stability of Structures & 1  & 55 \\
            & Safety of Use & 1  & 9 \\
            & Health (Indoor Climate; Water and sewerage; and Humidity) & 3  & 16 \\
            & Acoustic Environment & 1  & 4 \\

            \hline
    Total & & 33 & 1688 \\        
    \bottomrule
\end{tabular}
\end{table}

\subsubsection*{Sentence Collection and Processing}~\label{sec:sent-collection}
\begin{figure}[hbt]
\centering
\centerline{
  \includegraphics[width=0.9\textwidth]{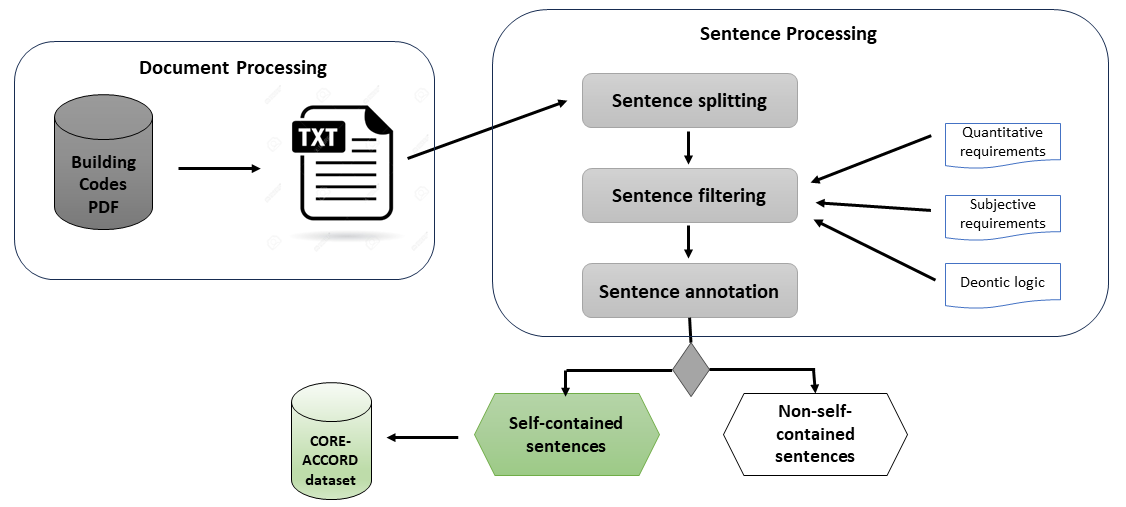}
}
\caption{\label{fig:dataPreparationMethodology} The Semi-Automatic CODE-ACCORD Data Preparation Methodology}
\end{figure}

We processed the data from the sources mentioned above to create our initial regulatory sentence collection, following the methodology illustrated in Figure \ref{fig:dataPreparationMethodology}. We excluded England's documents C, D, H and J from further processing due to the complexities associated with their format. Initially, all PDF documents sourced from both English and Finnish data repositories were converted into plain text. Following this, a semi-automated process was applied to filter regulatory sentences. During the filtering, we particularly focused on data that encompasses quantitative, subjective, and deontic requirements, essential for rule identification. Afterwards, the extracted sentences were manually filtered to select self-contained sentences that clearly expressed rules without relying on preceding or subsequent sentences. Each of the steps is further described below.

\paragraph{Document Processing: } This step consists of (i) extracting the textual data from the original PDF format using the PDFMiner library (\url{https://pypi.org/project/pdfminer/}), (ii) parsing the digital regulatory documents by performing actions like de-hyphenation, removing line breaks and footnotes, and (iii) removing the non-convertible tables and figures, and unnecessary sections.

\paragraph{Sentence Splitting:} This step consists of splitting the regulatory text into individual sentences by analysing word sequences and punctuation marks to determine sentence boundaries. To accomplish this, we used the sentence tokeniser provided by the Natural Language Toolkit (NLTK) library (\url{https://www.nltk.org/}).

\paragraph{Sentence Filtering:} This step initially involved an automatic filtering process to select sentences that pertain to regulations based on three distinct features described below. The filtering was performed using a keyword-matching approach.
\begin{itemize}[nosep]
    \item {\bf Quantitative requirements} refer to specific stipulations that can be expressed numerically or with quantitative terms. These requirements often specify precise values, measurements, or numerical criteria that must be met to ensure compliance with the regulations. Examples of quantitative requirements may include keywords such as {\emph {``less than''}}, \emph{``greater than''}, \emph{``equal''}, \emph{``at least''}, \emph{``higher than''}, \emph{``more than''}, \emph{``lower than''}, followed by numerical values or thresholds. The quantitative requirements are considered since they are mostly used in building codes for describing requirements~\cite{zhou2022integrating}.

    \item {\bf Subjective requirements} are stipulations that involve the use of subjective language or expressions. These requirements are not defined by precise numerical values or measurements, but rather by language that conveys recommendations, preferences, or suggestions. Subjective requirements often include terms like \emph{``should be''}, \emph{``recommended''}, \emph{``preferred''} or \emph{``advisable''}. While subjective in nature, these requirements are important in building regulations as they allow for flexibility and adaptation to different situations while still providing a framework for best practices and quality standards. To the best of our knowledge, existing research in the field of applying NLP for the automation of building regulations has not addressed subjective requirements in their analyses, methodologies, or datasets~\cite{zhou2022integrating}.

    \item {\bf Deontic logic} pertains to the logic that deals with the expression of permissions, obligations, prohibitions, and other normative statements. It is used to represent rules and requirements that are binding or mandatory, such as rules that specify what \emph{``must''}, \emph{``shall''}, \emph{``could''} or \emph{``prohibit''} within building regulations. Deontic logic plays a crucial role in modelling the normative aspects of these regulations, providing a formal framework to represent and reason about mandatory and discretionary requirements. Similarly to the subjective requirements, deontic logic has not been extensively considered in previous research efforts. This is primarily due to the focus of most research on quantitative requirements, given their higher frequency within building regulations~\cite{zhou2022integrating}.
\end{itemize} 

\noindent Following the automatic filtering, the next step involved manual curation to ensure the accuracy of the filtered sentences. One of the authors manually reviewed and removed false positives from the automatically filtered (auto-filtered) sentences, resulting in the semi-filtered sentence set, which was then used in subsequent steps.

\paragraph{Sentence Annotation: } The final step included manual annotation of filtered sentences to identify self-contained sentences, which are defined as sentences that encompass all necessary details without any linguistic co-references that cannot be resolved within the sentence itself. Moreover, self-contained sentences should not include references to external sources, such as sections, chapters, or documents, and should avoid the inclusion of incomplete or ambiguous concepts. Our work specifically focused on self-contained sentences because they convey rules that can be directly extracted without requiring cross-referencing or additional context that needs to be resolved by a human. This allows their information extraction and machine-readable rule formalisation to be fully automated. Non-self-contained sentences were excluded from the final regulatory sentence collection.

\subsubsection*{Sentence Statistics}

After applying the semi-automatic data collection methodology to the selected data sources, we obtained some noteworthy statistics summarised in Table \ref{tab:DataStats}. The total number of sentences was $20,674$, out of which $5,043$ were subjected to auto-filtering for capturing regulatory sentences, representing $24\%$ of the total sentences in the building regulations textual content. However, during the curation process of auto-filtered sentences, an unintended consequence of the sentence-splitting approach was noticed in a few documents due to conversion errors caused by PDF formatting. Some sentences were found to be grouped together with section and subsection headers, and introductory sentences that often contain colons and semi-colons, resulting in incorrect sentence structures. To address this, manual tweaking was carried out during sentence filtering to extract the sentences carefully. The combined sentences were cross-checked against the original documents, and additional elements were removed. Following this step, the false positives in auto-filtered sentences were also removed resulting in 4,459 semi-filtered sentences. From these, 1,246 self-contained sentences were manually extracted for the annotation task described in the following section.

\begin{table}[hbt!]
\caption{Statistics of the outputs of CODE-ACCORD semi-automatic data preparation methodology}\label{tab:DataStats}%
\begin{tabular}{@{}lrrrr@{}}
\toprule
Regulations & \#Sentences  & \#Auto-filtered Sentences & \#Semi-filtered Sentences & \#Self-contained Sentences \\
\midrule
England  & $19201$  & $4219$ & $3695$ & $963$   \\
    
    Finland & $1473$ & $824$ & $764$ & $283$ \\
            \hline
    Total & 20674 & 5043 & 4459 & 1246 \\        
    \bottomrule
\end{tabular}
\end{table}

\subsection*{Data Annotation Methodology} \label{sec:data-annotation}

In data annotation, we primarily focused on extracting information from text to facilitate automatic rule generation. There are two key types of information found in the text: named entities and relations, which are essential for comprehending the ideas conveyed in natural language \cite{Jurafsky2009speech}. Hence, our primary focus in this work was on annotating entities and relations.

For CODE-ACCORD manual annotations, we used a group of 12 annotators with either a computer science or a civil engineering/construction background. Since this work targets the automation of compliance checking using machine learning concepts, we believe it is important to involve experts from both areas in the annotation process. All the annotators are from the ACCORD project (\url{https://accordproject.eu/}) and received allocated compensation for the work packages involved in this task. We used the LightTag text annotation platform \cite{Perry2021lighttag} to collect human annotations, considering its coverage of different text annotations, including entities and relations, project management support and user-friendly interfaces. We provide an overview of our annotation methodology, highlighting the key concepts below. More detailed information is available in our comprehensive annotation manual (\url{https://github.com/Accord-Project/CODE-ACCORD/blob/main/annotated_data/Annotation_Strategy_V1.0.0.pdf}). Further details on annotation quality are provided in the Technical Validation section.

\subsubsection*{Entity Annotation} \label{sec:entity-annotation}

By named entity/entity, we refer to a specific piece of information or a concept that can be categorised. Simply, named entities are anything that can be referred to using a proper name \cite{Jurafsky2009speech}. 

\paragraph{\textbf{Entity Categories:}} Following the idea proposed in \cite{zhang2021deep, zhou2022integrating}, we picked four named entity categories, described in Table \ref{tab:ne-labels}, for entity annotation. However, deviating from previous work, we adopted a simple category structure, mainly aiming at the generalisability of our annotation approach across different subdomains, such as structure, fire safety and accessibility, when defining the named entities. Also, we considered the coverage of all information in a regulatory sentence.  

\begin{table}[hbt!]
\caption{Entity categories} 
\label{tab:ne-labels}%
\begin{tabular}{p{0.15\textwidth}p{0.76\textwidth}}
\toprule
Named Entity & Description\\
\midrule
\textcolor{blue}{object} & An ontological concept which represents a thing that is subject to a particular requirement (e.g. window, fire door) \\  
\textcolor{orange}{property} & Property of an object (e.g., width, height) \\
\textcolor{gray}{quality} & Quality or uncountable characteristic of an object/property (e.g. horizontal, self-closing) \\
\textcolor{amber}{value} & A standard or a numerical value that defines a quantity (e.g. 1,500 millimetres, five per cent) \\
\bottomrule
\end{tabular}
\end{table}

\paragraph{\textbf{Annotation Process:}} The entity annotation process mainly consisted of two steps: (1) mark entity text spans and (2) assign entity categories. A few annotated samples are shown in Table \ref{tab:ne-samples}. As can be seen, the selected categories are versatile enough to capture all entities in different sentence structures. Also, these samples are from Accessibility and Fire Safety regulations to indicate the general applicability of our annotation strategy in different subdomains.

\begin{table}[hbt!]
\caption{Sample named entity annotations} \label{tab:ne-samples}%
\begin{tabular}{p{0.4\textwidth}p{0.52\textwidth}}
\toprule
Sentence & Annotated Sentence\\
\midrule
The gradient of the passageway located in an outdoor space may not exceed five per cent. &
The \textcolor{orange}{<property>gradient</property>} of the \textcolor{blue}{<object>passageway</object>} located in an \textcolor{blue}{<object>outdoor space</object>} may not exceed \textcolor{amber}{<value>five per cent</value>}. \\
There shall be a horizontal landing with a length of at least 1,500 millimetres at the lower and upper end of the ramp. & 
There shall be a \textcolor{gray}{<quality>horizontal</quality>} \textcolor{blue}{<object>landing</object>} with a \textcolor{orange}{<property>length</property>} of at least \textcolor{amber}{<value>1,500 millimetres</value>} at the \textcolor{orange}{<property>lower and upper end</property>} of the \textcolor{blue}{<object>ramp</object>}. \\
A fire door must be self-closing and self-bolting. & A \textcolor{blue}{<object>fire door</object>} must be \textcolor{gray}{<quality>self-closing</quality>} and \textcolor{gray}{<quality>self-bolting</quality>}. \\

\bottomrule
\end{tabular}
\end{table}

\subsubsection*{Relation Annotation} \label{sec:relation-annotation}

Relations are semantic connections/associations among entities in the text \cite{Jurafsky2009speech}. Extraction of relations together with entities is a crucial process to transform information embedded in unstructured texts into structured data formats such as knowledge graphs. 

\paragraph{\textbf{Relation Categories:}}
Due to the lack of generic, non-domain-specific relation annotation approaches in compliance checking, after carefully analysing the possible relations in the regulatory text, we identified ten relation categories, detailed in Table \ref{tab:re-labels}. Similar to our approach in defining entity labels, we mainly focused on the generalisability across different subdomains and coverage of semantic information when identifying the relation categories. The final category, \textit{`none'}, is added considering the potential model requirements for identifying instances without relation between entity pairs.

\begin{table}[hbt!]
\caption{Relation categories} 
\label{tab:re-labels}%
\begin{tabular}{p{0.15\textwidth}p{0.76\textwidth}}
\toprule
Relation & Description\\
\midrule
selection  & A limit to the scope of an object/property based on another object or a quality \\  
necessity  & A qualitative/subjective/existential necessity of an object/property (e.g., should, should have, shall be, etc.)  \\
part-of  & Being a part of an object/property  \\
not-part-of  & Not being a part of an object/property  \\
greater & A value that should be greater than to \\
greater-equal & A value that should be greater than or equal to \\
equal & A value that should be equal to \\
less-equal & A value that should be less than or equal to \\
less & A value that should be less than to \\
none & No relation \\
\bottomrule
\end{tabular}
\end{table}

\begin{table}[H]
\caption{Sample relation annotations} \label{tab:re-samples}
\vspace{-1em}%
\begin{tabular}{p{0.2\textwidth}p{0.64\textwidth}p{0.08\textwidth}}
\toprule
Sentence & Entity Pair & Relation\\
\midrule
\multirow{3}{*}{\parbox{0.2\textwidth}{The gradient of the passageway located in an outdoor space may not exceed five per cent.}} &
The \textcolor{orange}{<property>gradient</property>} of the \textcolor{blue}{<object>passageway</object>} located in an outdoor space may not exceed five per cent. & part-of \\
\cmidrule{2-3}
& The gradientof the \textcolor{blue}{<object>passageway</object>} located in an \textcolor{blue}{<object>outdoor space</object>} may not exceed five per cent. & part-of \\
\cmidrule{2-3}
& The \textcolor{orange}{<property>gradient</property>} of the passageway located in an outdoor space may not exceed \textcolor{amber}{<value>five per cent</value>}. & less-equal \\
\midrule
\multicolumn{2}{l}{Entity-relation representation} &\\
& & \includegraphics[width=0.67\textwidth, right]{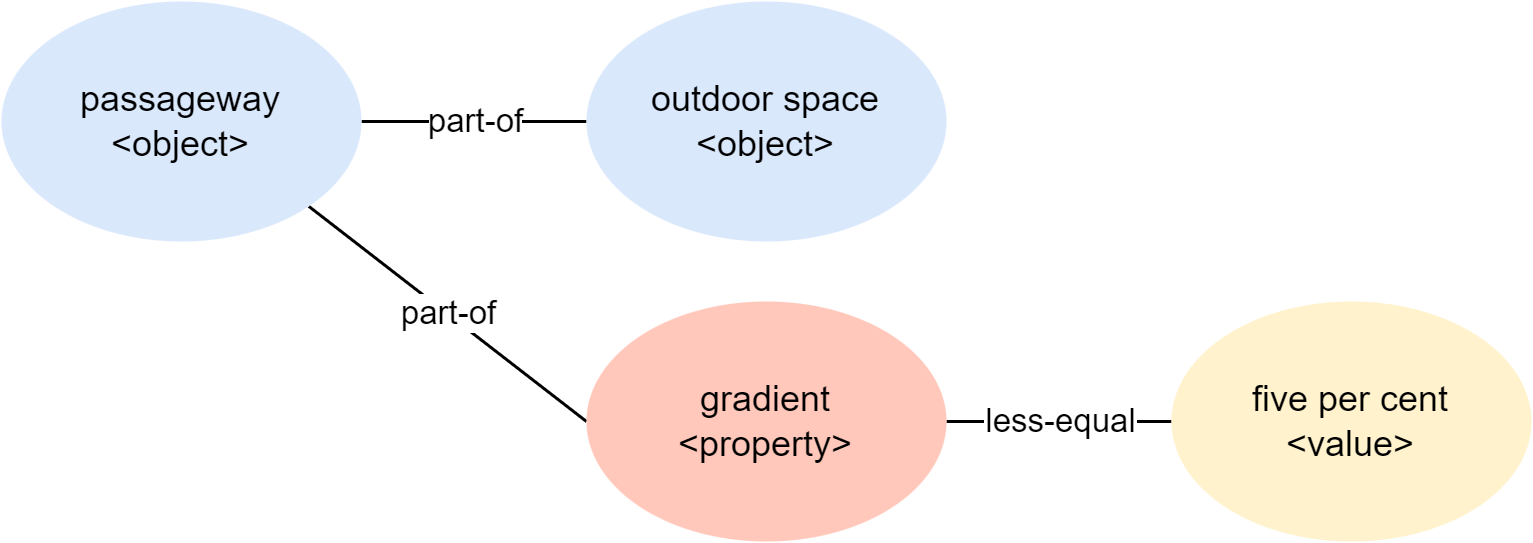}\\

\midrule

\multirow{3}{*}{\parbox{0.2\textwidth}{There shall be a horizontal landing with a length of at least 1,500 millimetres at the lower and upper end of the ramp.}} & 
There shall be a \textcolor{gray}{<quality>horizontal</quality>} \textcolor{blue}{<object>landing</object>} with a length of at least 1,500 millimetres at the lower and upper end of the ramp. & selection \\
\cmidrule{2-3}
& There shall be a horizontal \textcolor{blue}{<object>landing</object>} with a \textcolor{orange}{<property>length</property>} of at least 1,500 millimetres at the lower and upper end of the ramp. & part-of \\
\cmidrule{2-3}
& There shall be a horizontal landing with a \textcolor{orange}{<property>length</property>} of at least \textcolor{amber}{<value>1,500 millimetres</value>} at the lower and upper end of the ramp. & greater-equal \\
\cmidrule{2-3}
& There shall be a horizontal \textcolor{blue}{<object>landing</object>} with a length of at least 1,500 millimetres at the \textcolor{orange}{<property>lower and upper end</property>} of the ramp. & necessity \\
\cmidrule{2-3}
& There shall be a horizontal landing with a length</property of at least 1,500 millimetres at the \textcolor{orange}{<property>lower and upper end</property>} of the \textcolor{blue}{<object>ramp</object>}. & part-of \\
\midrule
\multicolumn{2}{l}{Entity-relation representation} & \\
& & \includegraphics[width=0.67\textwidth, right]{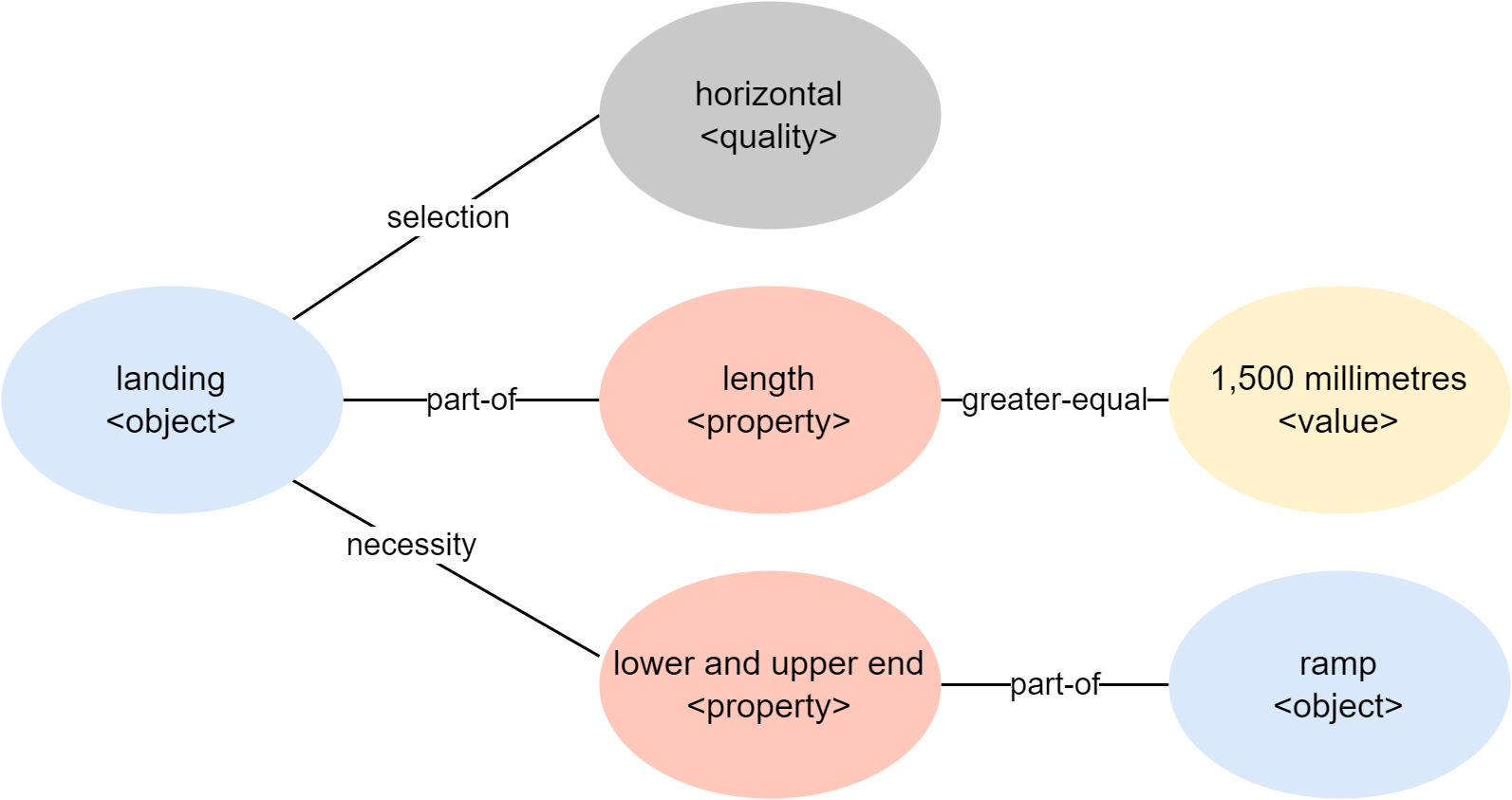}\\

\bottomrule
\end{tabular}
\end{table}

\paragraph{\textbf{Annotation Process:}} Overall, the relation annotation process consisted of four steps: (1) mark entity text spans, (2) assign entity categories, (3) identify entity pairs which form relations, and (4) assign relation categories. This task has proven more challenging than entity annotations, primarily due to its multiple intricate steps and the potential for error propagation. However, we instructed the annotators to adhere to the entire flow and extracted entity and relation annotations upon completing the second and fourth steps, allowing them the flexibility to highlight all relevant content simultaneously. Furthermore, this approach enabled them to review the provided annotations by examining the final entity-relation representation. We only applied the manual annotation process targeting the first nine entity categories without the \textit{`none'} category because once all the available relations are known, the remaining possible entity pairs form the no relations. Table \ref{tab:re-samples} shows a few annotated samples following the complete process.

\section*{Data Records}

The CODE-ACCORD dataset is available at \url{https://doi.org/10.5281/zenodo.10210022}\cite{codeaccord}. The data repository contains three main folders: \textit{English{\textunderscore}Regulations}, \textit{Finnish{\textunderscore}Regulations} and \textit{Annotated{\textunderscore}Data}. The first two folders include the textual data utilised for sentence collection, and the latter contains the annotated data, as further described below. 

\subsection*{Sentence Collection}
There are separate data folders, named  \textit{English{\textunderscore}Regulations} and \textit{Finnish{\textunderscore}Regulations} for England's and Finland's regulatory text, respectively. Each folder's hierarchy is structured as follows. Within the main data folder, there are two primary sub-folders. The first sub-folder, \textit{PDF}, contains the original PDF files of the regulations documents. The second sub-folder, \textit{Text and CSV}, is where TXT and CSV versions of the PDF files are stored after undergoing various pre-processing stages. This \textit{Text and CSV} sub-folder consists of eight sub-folders, each corresponding to a specific pre-processing step. They are meticulously organised in sequential order to facilitate systematic data handling. The first sub-folder, \textit{RawTextData}, contains the initial raw text data obtained through PDF-to-text conversion. The subsequent sub-folder, \textit{CleanedData-RawText}, holds the cleaned data derived from the initial raw text. \textit{AllSentences} sub-folder contains all sentences extracted from the cleaned text. Next, \textit{AutoFilteredSentences} comprises sentences that have been automatically filtered, following the specific criteria described in the Data Collection Methodology above. \textit{ManuallyFilteredSentences} contains manually curated sentences to ensure consistency and remove unnecessary content. The \textit{FinalFilteredSentences} sub-folder stores the ultimate raw text of the semi-automatically filtered sentences after eliminating empty lines and redundant information. Moving on, \textit{CSV-FinalFilteredSentences} presents the sentences from the \textit{FinalFilteredSentences} folder in CSV format, preparing the data for the final sub-folder, \textit{Classification} which categorises the sentences as either \emph{``self-contained''} or  \emph{``others''}, where only the self-contained sentences were considered for final data annotation. This structured hierarchy streamlines the data processing and analysis of regulations documents, ensuring an organised and efficient workflow.

\subsection*{Annotated Data} \label{sec:data-formats}

A randomly selected 862 sentences out of 1246 self-contained sentences extracted from the building regulations of England and Finland underwent our comprehensive data annotation process, targeting both entities and relations, which are essential for extracting information from text. All annotated data are available in \textit{Annotated{\textunderscore}Data} folder. There are two sub-folders named \textit{entities} and \textit{relations} within it, which hold entity-annotated data and relation-annotated data, respectively. Each folder has three CSV files named \textit{all.csv}, \textit{train.csv} and \textit{test.csv}. File \textit{all.csv} contains the full dataset. \textit{train.csv} has 80\% of the full dataset, which can be used to train machine learning models, and \textit{test.csv} has the remaining 20\%, which can be used for models' performance testing. All three files within one folder follow the same format. The entity and relation data file formats are further described below.

\subsubsection*{Entity Data Format} \label{sec:entity-data-format}

\begin{figure}[!hbt]
\centering
\centerline{
  \includegraphics[width=0.55\textwidth]{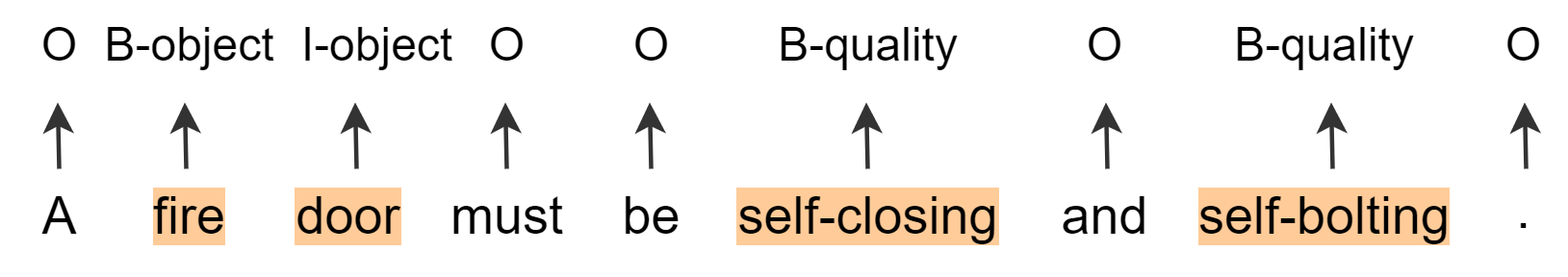}
}
\caption{Sample of entity labels in BIO format}
\label{fig:bio-sample}
\end{figure}

The format of an entity data file available within the \textit{ner} folder is summarised in Table \ref{tab:file-format-entity}. The entity annotations are available in the BIO (Beginning, Inside, Outside) format, which is considered the standard for information extraction tasks \cite{ramshaw1995text}, as shown in the sample in Figure \ref{fig:bio-sample}. \textit{object} and \textit{quality} tags shown in the figure are two categories of the selected entity categories, which are further described in the Data Annotation Methodology above.

\begin{table}[hbt!]
\caption{Format of entity data file} 
\label{tab:file-format-entity}%
\begin{tabular}{p{0.15\textwidth}p{0.76\textwidth}}
\toprule
Attribute & Description\\
\midrule
example{\textunderscore}id &  Unique ID assigned for each sentence\\  
content & Original textual content of the sentence\\
processed{\textunderscore}content &  Tokenised (using NLTK's word{\textunderscore}tokenize package) textual content of the sentence\\
label & Entity labelled sequence in IOB format\\
metadata & Additional information of sentence (i.e. original approved document from which the sentence is extracted)\\
\bottomrule
\end{tabular}
\end{table}

\subsubsection*{Relation Data Format} \label{sec:relation-data-format}
The format of a relation data file available within the \textit{re} folder is summarised in Table \ref{tab:file-format-relation}. We adopted the following format to tag the entity pair denoting the relationship within a data sample, in accordance with formats utilised in recent studies \cite{bastos2021recon, plum2022biographical}:
\begin{tcolorbox}
The <e1>gradient</e1> of the <e2>passageway</e2> located in an outdoor space may not exceed five per cent. 
\end{tcolorbox}
\noindent The special tags <e1> and </e1> represent the start and end of the first entity that appeared in the sentence. Similarly, <e2> and </e2> represent the second entity.

\begin{table}[hbt!]
\caption{Format of relation data file} 
\label{tab:file-format-relation}%
\begin{tabular}{p{0.15\textwidth}p{0.76\textwidth}}
\toprule
Attribute & Description\\
\midrule
example{\textunderscore}id &  Unique ID assigned for each sentence\\  
content & Original textual content of the sentence\\
metadata & Additional information of sentence (i.e. original approved document from which the sentence is extracted)\\
tagged{\textunderscore}sentence & Sentence with tagged entity pair \\
relation{\textunderscore}type & Category of the relation in between the tagged entity pair \\
\bottomrule
\end{tabular}
\end{table}

\subsection*{Descriptive Statistics}

The statistical analysis of the final annotated dataset is vital for future efforts that will use CODE-ACCORD as a resource for automating the conversion of textual rules into machine-readable formats, facilitating Automated Compliance Checking (ACC). 

\vspace{1em}
\begin{minipage}{.46\textwidth}
\centering
\includegraphics[width=0.96\textwidth]{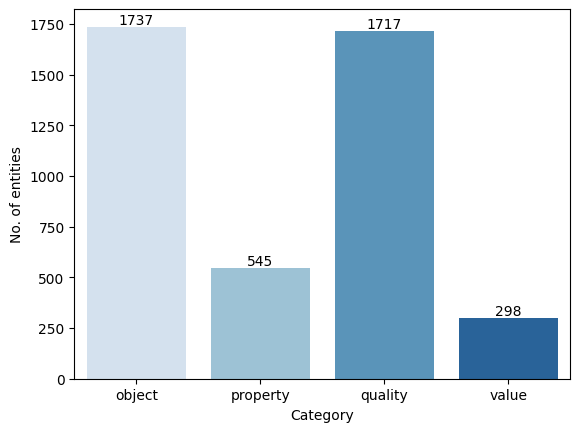}
\captionof{figure}{Distribution of entity categories}
\label{fig:cat-entity}            
\end{minipage}
\noindent\begin{minipage}{.44\textwidth}
\centering
\includegraphics[width=0.96\textwidth]
{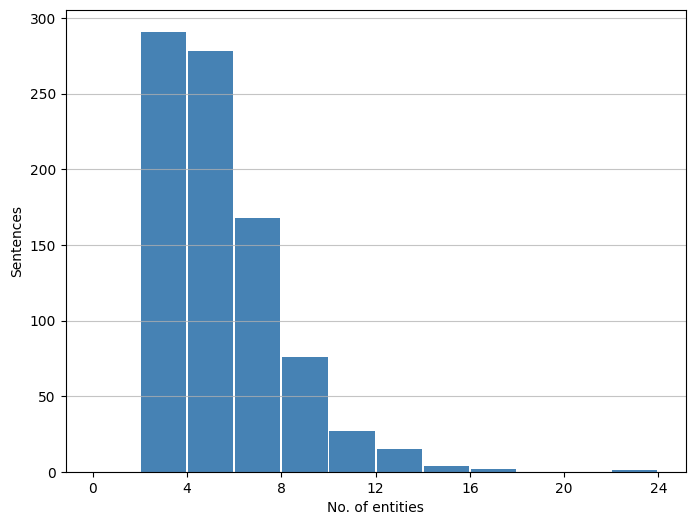}
\captionof{figure}{Distribution of the number of entities per sentence}
\label{fig:histo-ner}           
\end{minipage}%

\begin{figure}[hbt]
\centering
\centerline{
  \includegraphics[width=0.9\textwidth]{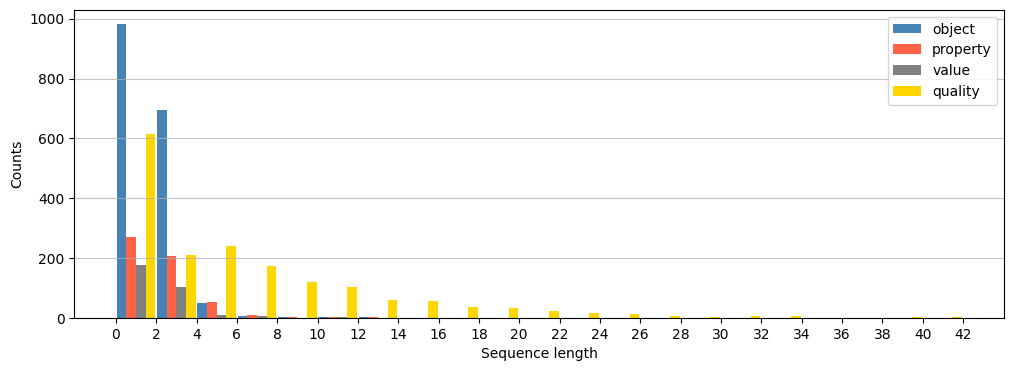}
}
\caption{Sequence length distribution of annotated text spans as entities}
\label{fig:histo-entity}
\end{figure}

\paragraph{\textbf{Entity Statistics:}} Our final entity annotated dataset contains 862 sentences. It has 4,297 entities distributed over four categories, as shown in Figure \ref{fig:cat-entity}. The illustrated distribution of the number of entities per sentence in Figure \ref{fig:histo-ner} provides a detailed insight into the annotated data. As can be seen, most sentences contain up to five entities. We further analysed the sequence lengths of text spans from each category, and the resulting histograms are presented in Figure \ref{fig:histo-entity}. Most of the entity spans are composed of one or two words/tokens. However, overall, there are more lengthy text spans under \textit{quality} than in the other categories.

\paragraph{\textbf{Relation Statistics:}} Altogether, we annotated 862 sentences, resulting in 3,329 relations. We automatically identified the unannotated entity pairs within sentences as unrelated entity pairs which belong to the \textit{`none'} category. Out of 8,104 samples categorised as  \textit{`none'}, we included a random subset of 1,000 in our final dataset to ensure a balanced distribution with other relations. The breakdown of a total of 4,329 relations across ten categories is depicted in Figure \ref{fig:cat-re}. Additionally, Figure \ref{fig:histo-re} illustrates the distribution of the number of relations per sentence. Most sentences contained two or three relations, although a minority had over ten relations. 

\vspace{1em}
\noindent\begin{minipage}{.56\textwidth}
\centering
\includegraphics[width=0.96\textwidth]{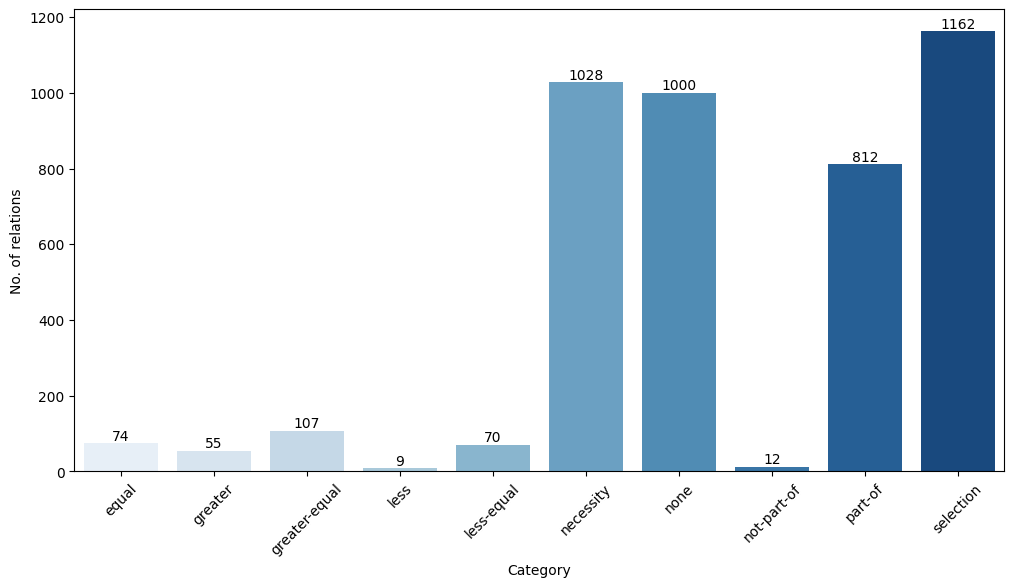}
\captionof{figure}{Distribution of relation categories}
\label{fig:cat-re}            
\end{minipage}%
\begin{minipage}{.44\textwidth}
\centering
\includegraphics[width=0.96\textwidth]{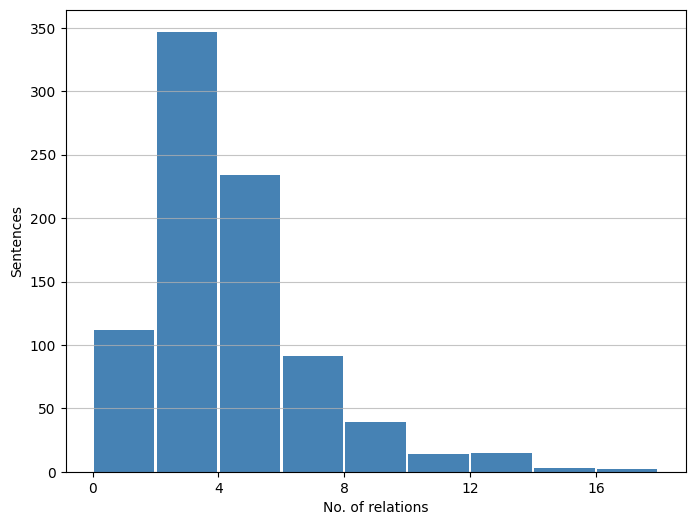}
\captionof{figure}{Distribution of the number of relations per sentence}
\label{fig:histo-re}            
\end{minipage}

\section*{Technical Validation}

\subsection*{Annotation Quality}

We employed several methods to ensure the quality of the annotations. The annotation strategy was initially designed based on two fundamental types of information in the text (i.e. entities and relations) along with available methods. It was then refined through multiple iterations, incorporating feedback from several domain experts. Additionally, we conducted two rounds of test annotations with a team of two annotators familiar with the strategy to validate its feasibility and coverage before finalising it.

We conducted the annotations for both entities and relations in multiple rounds. The first round served as a training session for the annotators. They were given sufficient time to review the annotation strategy and examples, with additional clarifications provided upon request before beginning the round. After this round, a question-and-answer session was held to address any queries with the help of several domain experts. The actual annotation rounds then began, with a total of seven rounds. Each sentence was independently annotated by two or three annotators through these rounds.

\begin{figure}[!hbt]
\centering
\centerline{
  \includegraphics[width=0.5\textwidth]{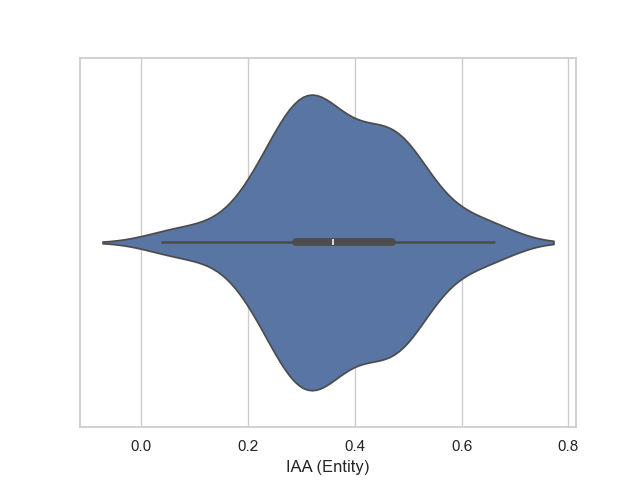}
}
\caption{Distribution of entity IAA values}
\label{fig:iaa-entity}
\end{figure}

To measure the quality of annotations, we calculated Inter Annotator Agreements (IAAs) throughout our rounds. As the entity IAA, we used the pairwise relative agreement of entities. The annotator A's agreement with Annotator B is calculated using Equation \ref{eq:iaa-entity} \cite{Perry2021lighttag}. An entity annotated by one annotator is considered a match to an entity by another annotator only if the marked text span and assigned label are equal. Figure \ref{fig:iaa-entity} summarises the distribution of entity IAA values obtained throughout the annotation rounds. The mean IAA is 0.37, with a maximum of 0.66. The task's complexities (i.e. two-step annotation process and domain-specific knowledge requirements) and strict matching criteria used during the agreement calculation can be identified as the primary factors contributing to this distribution. However, given the further complexities associated with relation annotations following its four-step process, we limited our IAA calculations only to entities.

\begin{equation} \label{eq:iaa-entity}
\text{A's agreement with B} = \frac{\text{Number of B's entities matched with A's entities}}{\text{Total entities annotated by B}}
\end{equation}

To enhance the accuracy of the annotations, each annotation round was followed by a curation round to determine the final annotations. Three members from the annotation group, who are domain experts, served as data curators. Their curation jobs were carefully assigned without any overlaps with the annotation jobs. During curation, the final annotations for all entities and relations with disagreements between annotators have been decided by the curator, considering the proposed annotations and the overall entity-relation representation of each sentence. Additionally, the content of each sentence was cross-checked against the final entity-relation representation or knowledge graph to confirm its completeness and to verify that all relevant information in the text was accurately captured.

\subsection*{Data Splits Quality}

\begin{figure}[!hbt]
\centering
\centerline{
  \includegraphics[width=\textwidth]{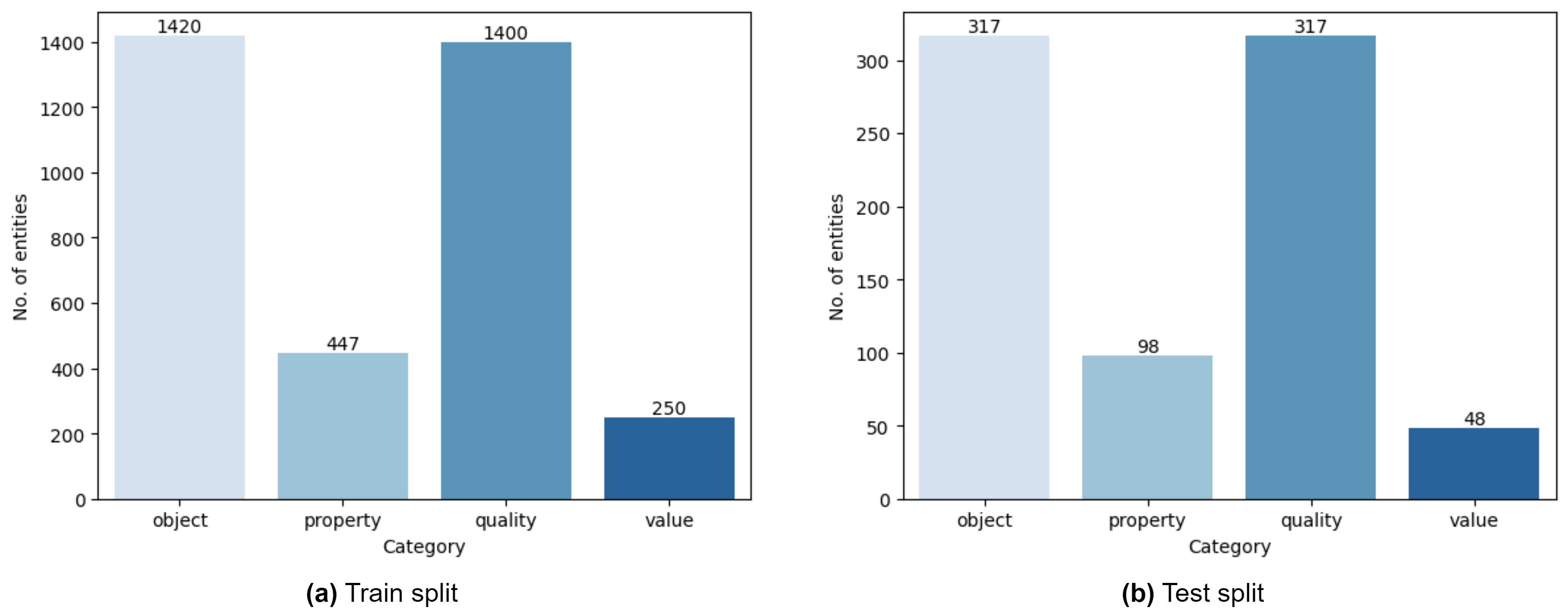}
}
\caption{Distribution of the entity categories in train and test data splits}
\label{fig:cat-ner-splits}
\end{figure}

Alongside the complete annotated dataset, CODE-ACCORD offers two data splits: train and test, for each annotation type (i.e. entities and relations). Each train split comprises 80\% of the corresponding full dataset and is intended for training machine learning models. The remaining 20\% forms the test splits, designed for evaluating model performance. It is essential to ensure an equivalent class distribution across data splits, as it directly affects the model training and evaluation procedures and the overall accuracy of the final model. Thus, we used stratified sampling to create these splits, ensuring that the class distribution in each split mirrors that of the original data. We further validated this fact using exploratory data analysis resulting in Figures \ref{fig:cat-ner-splits} and \ref{fig:cat-re-splits}, which illustrate the distribution of entity and relation categories in the train and test data splits, respectively, demonstrating the consistency of distributions across these splits.


\textbf{\begin{figure}[!hbt]
\centering
\centerline{
  \includegraphics[width=\textwidth]{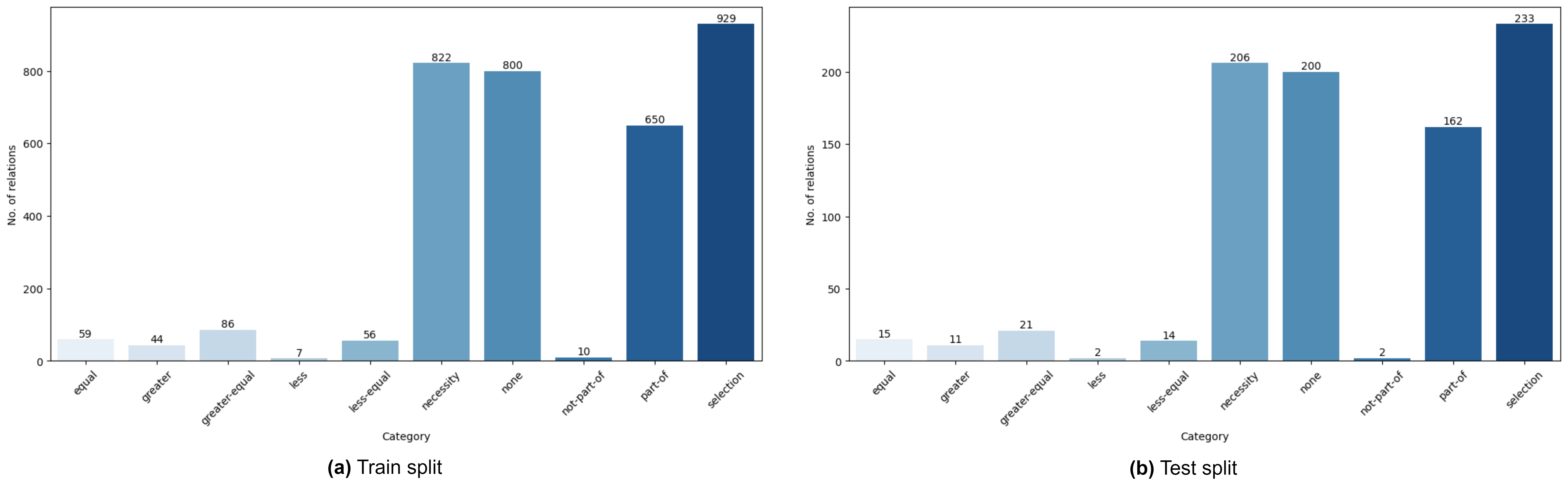}
}
\caption{Distribution of the relation categories in train and test data splits}
\label{fig:cat-re-splits}
\end{figure}}




\section*{Code availability}


\textcolor{blue}{CODE-ACCORD GitHub repository is available at \url{https://github.com/Accord-Project/CODE-ACCORD/}.}


\bibliography{sample}

\begin{thebibliography}{10}
\urlstyle{rm}
\expandafter\ifx\csname url\endcsname\relax
  \def\url#1{\texttt{#1}}\fi
\expandafter\ifx\csname urlprefix\endcsname\relax\def\urlprefix{URL }\fi
\expandafter\ifx\csname doiprefix\endcsname\relax\def\doiprefix{DOI: }\fi
\providecommand{\bibinfo}[2]{#2}
\providecommand{\eprint}[2][]{\url{#2}}

\bibitem{fuchs2021natural}
\bibinfo{author}{Fuchs, S.} \& \bibinfo{author}{Amor, R.}
\newblock \bibinfo{title}{Natural language processing for building code interpretation: A systematic literature review}.
\newblock In \emph{\bibinfo{booktitle}{Proceedings of the Conference CIB W78}}, vol. \bibinfo{volume}{2021}, \bibinfo{pages}{11--15} (\bibinfo{year}{2021}).

\bibitem{zhang2023unpacking}
\bibinfo{author}{Zhang, Z.}, \bibinfo{author}{Ma, L.} \& \bibinfo{author}{Nisbet, N.}
\newblock \bibinfo{journal}{\bibinfo{title}{Unpacking ambiguity in building requirements to support automated compliance checking}}.
\newblock {\emph{\JournalTitle{Journal of Management in Engineering}}} \textbf{\bibinfo{volume}{39}}, \bibinfo{pages}{04023033}, \url{https://doi.org/10.1061/JMENEA.MEENG-5359} (\bibinfo{year}{2023}).
\newblock \eprint{https://ascelibrary.org/doi/pdf/10.1061/JMENEA.MEENG-5359}.

\bibitem{MALSANE201551}
\bibinfo{author}{Malsane, S.}, \bibinfo{author}{Matthews, J.}, \bibinfo{author}{Lockley, S.}, \bibinfo{author}{Love, P.~E.} \& \bibinfo{author}{Greenwood, D.}
\newblock \bibinfo{journal}{\bibinfo{title}{Development of an object model for automated compliance checking}}.
\newblock {\emph{\JournalTitle{Automation in Construction}}} \textbf{\bibinfo{volume}{49}}, \bibinfo{pages}{51--58}, \url{https://doi.org/10.1016/j.autcon.2014.10.004} (\bibinfo{year}{2015}).

\bibitem{zhou2022integrating}
\bibinfo{author}{Zhou, Y.-C.}, \bibinfo{author}{Zheng, Z.}, \bibinfo{author}{Lin, J.-R.} \& \bibinfo{author}{Lu, X.-Z.}
\newblock \bibinfo{journal}{\bibinfo{title}{Integrating nlp and context-free grammar for complex rule interpretation towards automated compliance checking}}.
\newblock {\emph{\JournalTitle{Comput. Ind.}}} \textbf{\bibinfo{volume}{142}}, \url{https://doi.org/10.1016/j.compind.2022.103746} (\bibinfo{year}{2022}).

\bibitem{Zhang2012}
\bibinfo{author}{Zhang, J.} \& \bibinfo{author}{El-Gohary, N.}
\newblock \bibinfo{title}{Extraction of construction regulatory requirements from textual documents using natural language processing techniques}.
\newblock In \emph{\bibinfo{booktitle}{Congress on Computing in Civil Engineering, Proceedings}}, \bibinfo{pages}{453--460}, \url{https://doi.org/10.1061/9780784412343.0057} (\bibinfo{year}{2012}).

\bibitem{Locatelli2021}
\bibinfo{author}{Locatelli, M.}, \bibinfo{author}{Seghezzi, E.}, \bibinfo{author}{Pellegrini, L.}, \bibinfo{author}{Tagliabue, L.~C.} \& \bibinfo{author}{Di~Giuda, G.~M.}
\newblock \bibinfo{journal}{\bibinfo{title}{Exploring natural language processing in construction and integration with building information modeling: A scientometric analysis}}.
\newblock {\emph{\JournalTitle{Buildings}}} \textbf{\bibinfo{volume}{11}}, \url{https://doi.org/10.3390/buildings11120583} (\bibinfo{year}{2021}).

\bibitem{hjelseth2011capturing}
\bibinfo{author}{Hjelseth, E.} \& \bibinfo{author}{Nisbet, N.}
\newblock \bibinfo{title}{Capturing normative constraints by use of the semantic mark-up rase methodology}.
\newblock In \emph{\bibinfo{booktitle}{Proceedings of CIB W78-W102 Conference}}, \bibinfo{pages}{1--10} (\bibinfo{year}{2011}).

\bibitem{Beach2015rule}
\bibinfo{author}{Beach, T.}, \bibinfo{author}{Rezgui, Y.}, \bibinfo{author}{Li, H.} \& \bibinfo{author}{Kasim, T.}
\newblock \bibinfo{journal}{\bibinfo{title}{A rule-based semantic approach for automated regulatory compliance in the construction sector}}.
\newblock {\emph{\JournalTitle{Expert Syst. Appl.}}} \textbf{\bibinfo{volume}{42}}, \bibinfo{pages}{5219–5231}, \url{https://doi.org/10.1016/j.eswa.2015.02.029} (\bibinfo{year}{2015}).

\bibitem{LEE201649}
\bibinfo{author}{Lee, H.}, \bibinfo{author}{Lee, J.-K.}, \bibinfo{author}{Park, S.} \& \bibinfo{author}{Kim, I.}
\newblock \bibinfo{journal}{\bibinfo{title}{Translating building legislation into a computer-executable format for evaluating building permit requirements}}.
\newblock {\emph{\JournalTitle{Automation in Construction}}} \textbf{\bibinfo{volume}{71}}, \bibinfo{pages}{49--61}, \url{https://doi.org/10.1016/j.autcon.2016.04.008} (\bibinfo{year}{2016}).
\newblock \bibinfo{note}{The Special Issue of 32nd International Symposium on Automation and Robotics in Construction}.

\bibitem{zhang2016semantic}
\bibinfo{author}{Zhang, J.} \& \bibinfo{author}{El-Gohary, N.~M.}
\newblock \bibinfo{journal}{\bibinfo{title}{Semantic nlp-based information extraction from construction regulatory documents for automated compliance checking}}.
\newblock {\emph{\JournalTitle{Journal of Computing in Civil Engineering}}} \textbf{\bibinfo{volume}{30}}, \bibinfo{pages}{04015014}, \url{https://doi.org/10.1061/(ASCE)CP.1943-5487.0000346} (\bibinfo{year}{2016}).
\newblock \eprint{https://ascelibrary.org/doi/pdf/10.1061/\%28ASCE\%29CP.1943-5487.0000346}.

\bibitem{zhang2017integrating}
\bibinfo{author}{Zhang, J.} \& \bibinfo{author}{El-Gohary, N.~M.}
\newblock \bibinfo{journal}{\bibinfo{title}{Integrating semantic nlp and logic reasoning into a unified system for fully-automated code checking}}.
\newblock {\emph{\JournalTitle{Automation in Construction}}} \textbf{\bibinfo{volume}{73}}, \bibinfo{pages}{45--57}, \url{https://doi.org/10.1016/j.autcon.2016.08.027} (\bibinfo{year}{2017}).

\bibitem{zhou2017ontology}
\bibinfo{author}{Zhou, P.} \& \bibinfo{author}{El-Gohary, N.}
\newblock \bibinfo{journal}{\bibinfo{title}{Ontology-based automated information extraction from building energy conservation codes}}.
\newblock {\emph{\JournalTitle{Automation in Construction}}} \textbf{\bibinfo{volume}{74}}, \bibinfo{pages}{103--117}, \url{https://doi.org/10.1016/j.autcon.2016.09.004} (\bibinfo{year}{2017}).

\bibitem{xue2020building}
\bibinfo{author}{Xue, X.} \& \bibinfo{author}{Zhang, J.}
\newblock \bibinfo{journal}{\bibinfo{title}{Building codes part-of-speech tagging performance improvement by error-driven transformational rules}}.
\newblock {\emph{\JournalTitle{Journal of Computing in Civil Engineering}}} \textbf{\bibinfo{volume}{34}}, \bibinfo{pages}{04020035}, \url{https://doi.org/10.1061/(ASCE)CP.1943-5487.0000917} (\bibinfo{year}{2020}).
\newblock \eprint{https://ascelibrary.org/doi/pdf/10.1061/\%28ASCE\%29CP.1943-5487.0000917}.

\bibitem{zheng2022knowledge}
\bibinfo{author}{Zheng, Z.}, \bibinfo{author}{Zhou, Y.-C.}, \bibinfo{author}{Lu, X.-Z.} \& \bibinfo{author}{Lin, J.-R.}
\newblock \bibinfo{journal}{\bibinfo{title}{Knowledge-informed semantic alignment and rule interpretation for automated compliance checking}}.
\newblock {\emph{\JournalTitle{Automation in Construction}}} \textbf{\bibinfo{volume}{142}}, \bibinfo{pages}{104524}, \url{https://doi.org/10.1016/j.autcon.2022.104524} (\bibinfo{year}{2022}).

\bibitem{zhang2021deep}
\bibinfo{author}{Zhang, R.} \& \bibinfo{author}{El-Gohary, N.}
\newblock \bibinfo{journal}{\bibinfo{title}{A deep neural network-based method for deep information extraction using transfer learning strategies to support automated compliance checking}}.
\newblock {\emph{\JournalTitle{Automation in Construction}}} \textbf{\bibinfo{volume}{132}}, \bibinfo{pages}{103834}, \url{https://doi.org/10.1016/j.autcon.2021.103834} (\bibinfo{year}{2021}).

\bibitem{shen2022parallel}
\bibinfo{author}{Shen, Y.} \emph{et~al.}
\newblock \bibinfo{title}{Parallel instance query network for named entity recognition}.
\newblock In \bibinfo{editor}{Muresan, S.}, \bibinfo{editor}{Nakov, P.} \& \bibinfo{editor}{Villavicencio, A.} (eds.) \emph{\bibinfo{booktitle}{Proceedings of the 60th Annual Meeting of the Association for Computational Linguistics (Volume 1: Long Papers)}}, \bibinfo{pages}{947--961}, \url{https://doi.org/10.18653/v1/2022.acl-long.67} (\bibinfo{publisher}{Association for Computational Linguistics}, \bibinfo{address}{Dublin, Ireland}, \bibinfo{year}{2022}).

\bibitem{shen2023diffusionner}
\bibinfo{author}{Shen, Y.} \emph{et~al.}
\newblock \bibinfo{title}{{D}iffusion{NER}: Boundary diffusion for named entity recognition}.
\newblock In \bibinfo{editor}{Rogers, A.}, \bibinfo{editor}{Boyd-Graber, J.} \& \bibinfo{editor}{Okazaki, N.} (eds.) \emph{\bibinfo{booktitle}{Proceedings of the 61st Annual Meeting of the Association for Computational Linguistics (Volume 1: Long Papers)}}, \bibinfo{pages}{3875--3890}, \url{https://doi.org/10.18653/v1/2023.acl-long.215} (\bibinfo{publisher}{Association for Computational Linguistics}, \bibinfo{address}{Toronto, Canada}, \bibinfo{year}{2023}).

\bibitem{plum2022biographical}
\bibinfo{author}{Plum, A.}, \bibinfo{author}{Ranasinghe, T.}, \bibinfo{author}{Jones, S.}, \bibinfo{author}{Orasan, C.} \& \bibinfo{author}{Mitkov, R.}
\newblock \bibinfo{title}{Biographical semi-supervised relation extraction dataset}.
\newblock In \emph{\bibinfo{booktitle}{Proceedings of the 45th International ACM SIGIR Conference on Research and Development in Information Retrieval}}, SIGIR '22, \bibinfo{pages}{3121–3130}, \url{https://doi.org/10.1145/3477495.3531742} (\bibinfo{publisher}{Association for Computing Machinery}, \bibinfo{address}{New York, NY, USA}, \bibinfo{year}{2022}).

\bibitem{yang2023histred}
\bibinfo{author}{Yang, S.}, \bibinfo{author}{Choi, M.}, \bibinfo{author}{Cho, Y.} \& \bibinfo{author}{Choo, J.}
\newblock \bibinfo{title}{{H}ist{RED}: A historical document-level relation extraction dataset}.
\newblock In \bibinfo{editor}{Rogers, A.}, \bibinfo{editor}{Boyd-Graber, J.} \& \bibinfo{editor}{Okazaki, N.} (eds.) \emph{\bibinfo{booktitle}{Proceedings of the 61st Annual Meeting of the Association for Computational Linguistics (Volume 1: Long Papers)}}, \bibinfo{pages}{3207--3224}, \url{https://doi.org/10.18653/v1/2023.acl-long.180} (\bibinfo{publisher}{Association for Computational Linguistics}, \bibinfo{address}{Toronto, Canada}, \bibinfo{year}{2023}).

\bibitem{wang2021deep}
\bibinfo{author}{Wang, X.} \& \bibinfo{author}{El-Gohary, N.}
\newblock \emph{\bibinfo{title}{Deep Learning-Based Named Entity Recognition from Construction Safety Regulations for Automated Field Compliance Checking}}, \bibinfo{pages}{164--171} (\bibinfo{year}{2021}).
\newblock \eprint{https://ascelibrary.org/doi/pdf/10.1061/9780784483893.021}.

\bibitem{zheng2022pre}
\bibinfo{author}{Zheng, Z.}, \bibinfo{author}{Lu, X.-Z.}, \bibinfo{author}{Chen, K.-Y.}, \bibinfo{author}{Zhou, Y.-C.} \& \bibinfo{author}{Lin, J.-R.}
\newblock \bibinfo{journal}{\bibinfo{title}{Pretrained domain-specific language model for natural language processing tasks in the aec domain}}.
\newblock {\emph{\JournalTitle{Computers in Industry}}} \textbf{\bibinfo{volume}{142}}, \bibinfo{pages}{103733}, \url{https://doi.org/10.1016/j.compind.2022.103733} (\bibinfo{year}{2022}).

\bibitem{wang2023deep}
\bibinfo{author}{Wang, X.} \& \bibinfo{author}{El-Gohary, N.}
\newblock \bibinfo{journal}{\bibinfo{title}{Deep learning-based relation extraction and knowledge graph-based representation of construction safety requirements}}.
\newblock {\emph{\JournalTitle{Automation in Construction}}} \textbf{\bibinfo{volume}{147}}, \bibinfo{pages}{104696}, \url{https://doi.org/10.1016/j.autcon.2022.104696} (\bibinfo{year}{2023}).

\bibitem{athan2013oasis}
\bibinfo{author}{Athan, T.} \emph{et~al.}
\newblock \bibinfo{title}{{OASIS LegalRuleML}}.
\newblock In \emph{\bibinfo{booktitle}{Proceedings of the Fourteenth International Conference on Artificial Intelligence and Law}}, ICAIL '13, \bibinfo{pages}{3–12}, \url{https://doi.org/10.1145/2514601.2514603} (\bibinfo{publisher}{Association for Computing Machinery}, \bibinfo{address}{New York, NY, USA}, \bibinfo{year}{2013}).

\bibitem{fuchs2022neural}
\bibinfo{author}{Fuchs, S.}, \bibinfo{author}{Witbrock, M.}, \bibinfo{author}{Dimyadi, J.} \& \bibinfo{author}{Amor, R.}
\newblock \bibinfo{title}{Neural semantic parsing of building regulations for compliance checking}.
\newblock In \emph{\bibinfo{booktitle}{IOP Conference Series: Earth and Environmental Science}}, vol. \bibinfo{volume}{1101}, \bibinfo{pages}{092022}, \url{https://doi.org/10.1088/1755-1315/1101/9/092022} (\bibinfo{organization}{IOP Publishing}, \bibinfo{year}{2022}).

\bibitem{recski2024brise}
\bibinfo{author}{Recski, G.}, \bibinfo{author}{Ikl{\'o}di, E.}, \bibinfo{author}{Lellmann, B.}, \bibinfo{author}{Kov{\'a}cs, {\'A}.} \& \bibinfo{author}{Hanbury, A.}
\newblock \bibinfo{journal}{\bibinfo{title}{{BRISE-Plandok}: A german legal corpus of building regulations}}.
\newblock {\emph{\JournalTitle{Language Resources and Evaluation}}} \bibinfo{pages}{1--40}, \url{https://doi.org/10.1007/s10579-024-09747-7} (\bibinfo{year}{2024}).

\bibitem{Jurafsky2009speech}
\bibinfo{author}{Jurafsky, D.} \& \bibinfo{author}{Martin, J.~H.}
\newblock \emph{\bibinfo{title}{Speech and Language Processing (2Nd Edition)}} (\bibinfo{publisher}{Prentice-Hall, Inc.}, \bibinfo{address}{Upper Saddle River, NJ, USA}, \bibinfo{year}{2009}).

\bibitem{Perry2021lighttag}
\bibinfo{author}{Perry, T.}
\newblock \bibinfo{title}{{L}ight{T}ag: Text annotation platform}.
\newblock In \emph{\bibinfo{booktitle}{Proceedings of the 2021 Conference on Empirical Methods in Natural Language Processing: System Demonstrations}}, \bibinfo{pages}{20--27} (\bibinfo{publisher}{Association for Computational Linguistics}, \bibinfo{address}{Online and Punta Cana, Dominican Republic}, \bibinfo{year}{2021}).

\bibitem{codeaccord}
\bibinfo{author}{Hettiarachchi, H.} \emph{et~al.}
\newblock \bibinfo{journal}{\bibinfo{title}{Accord-project/{CODE-ACCORD}: v1.0.0}}.
\newblock {\emph{\JournalTitle{Zenodo}}} \url{https://doi.org/10.5281/zenodo.10210022} (\bibinfo{year}{2023}).

\bibitem{ramshaw1995text}
\bibinfo{author}{Ramshaw, L.} \& \bibinfo{author}{Marcus, M.}
\newblock \bibinfo{title}{Text chunking using transformation-based learning}.
\newblock In \emph{\bibinfo{booktitle}{Third Workshop on Very Large Corpora}} (\bibinfo{year}{1995}).

\bibitem{bastos2021recon}
\bibinfo{author}{Bastos, A.} \emph{et~al.}
\newblock \bibinfo{title}{{RECON}: Relation extraction using knowledge graph context in a graph neural network}.
\newblock In \emph{\bibinfo{booktitle}{Proceedings of the Web Conference 2021}}, WWW '21, \bibinfo{pages}{1673–1685}, \url{https://doi.org/10.1145/3442381.3449917} (\bibinfo{publisher}{Association for Computing Machinery}, \bibinfo{address}{New York, NY, USA}, \bibinfo{year}{2021}).

\end{thebibliography}

\section*{Acknowledgements}

This work is funded by the European Union’s Horizon Europe research and innovation programme under grant agreement no 101056973 (ACCORD). UK Participants in Horizon Europe Project [ACCORD] are supported by UKRI grant numbers [10040207] (Cardiff University), [10038999 ] (Birmingham City University) and [10049977] (Building Smart International).

\section*{Author contributions statement}


H.H.: Conceptualisation, Data preparation, Data annotation methodology, Data annotation, Data curation, Validation, Visualisation, Project administration, Writing - Original Draft, Writing - Review \& Editing.
A.D.: Conceptualisation, Data preparation methodology, Data preparation, Formal analysis, Data annotation methodology, Data annotation. Data curation, Writing - Original Draft. M.M.G.: Conceptualisation, Data annotation methodology, Supervision, Writing - Review \& Editing.
P.P.: Data annotation, Data curation, Writing - Review \& Editing.
N.B.: Data annotation, Writing - Review \& Editing.
K.B.: Data annotation, Writing - Review \& Editing.
G.C.: Data annotation, Writing - Review \& Editing.
M.H.: Data annotation, Writing - Review \& Editing.
M.J-M.: Data annotation, Writing - Review \& Editing.
T.M.: Data annotation, Writing - Review \& Editing.
S.P.: Data annotation, Writing - Review \& Editing.
H.T.: Data annotation, Writing - Review \& Editing.
A-R.H.T.: Data annotation, Writing - Review \& Editing.
E.V.: Conceptualisation, Data annotation methodology, Project administration, Funding acquisition, Writing - Review \& Editing.

\section*{Competing interests} 


The authors declare no competing interests.

\end{document}